# Messaging to Extra-Terrestrial Intelligence

[Alexander Zaitsev](#), IRE, Russia

*"It is more blessed to give than receive (Acts 20:35)."*

Those who propose, or oppose, sending Messages to Extra-Terrestrial Intelligence (METI) must contemplate the Hamlet-like question: "To send or not to send?" The science known as SETI deals with searching for messages *from* aliens. METI science deals with the creation of messages *to* aliens. Thus, SETI and METI proponents have quite different perspectives. SETI scientists are in a position to address only the local question "does Active SETI make sense?" In other words, would it be reasonable, for SETI success, to transmit with the object of attracting ETI's attention? In contrast to Active SETI, METI pursues not a local, but a more global purpose – to overcome the Great Silence in the Universe, bringing to our extraterrestrial neighbors the long-expected annunciation "You are not alone!" Thus, it follows that in the context of METI, the answer to the general question of transmissions from Earth requires competence beyond the membership of the highly specialized SETI Permanent Study Group of the International Academy of Astronautics (IAA-SPSG). We therefore propose that, for solution to the various current METI problems, we establish both a **METI Institute,** and METI Permanent Study Group within the IAA (an IAA-MPSG).

The respected SETI Institute has identified the following 7-dimensional space of unknown quantities for SETI consideration (see, for example: Jill Tarter, 1986. The Cosmic Haystack and Recent US SETI Programs):

1. Where to search?
2. When to search?

3. At what wavelength?
4. Type of polarization?
5. Power of radiation received?
6. How to demodulate the detected signals?
7. How to decode the received information?

This list can be adapted to aid in decisions regarding transmissions from Earth of our own radio messages to possible extraterrestrial civilizations. Transmission of interstellar radio messages (IRMs) is essentially a new kind of human activity, involving radiation of coherent signals from the Earth into space, addressed to other reasoning beings. Humans have always peered at the sky, in the hope of finding there intelligences beyond our own. METI thus implies a special and purposeful transmission. We can thus replace the terms connected with a search for radio signals, with terms associated with the transmission of same. In a more general treatment, a transformation from SETI to METI can occur as a transition from the science of merely separating those messages that already exist in the nature from artificial ones – namely Their reasonable radio signals – to the art of creating messages that do not exist in nature – namely our deliberate radio signals directed toward Them.

It would seem that there are two more new measurements in METI-space than there were in the case of SETI. Thus, the METI search space is a 9-dimensional one. We are compelled to consider such questions as "Why is it necessary to transmit and what we shall gain from doing so?" and "Is it dangerous to transmit messages to ETI?". In view of these two additional questions, we suggest that the proposed **METI Institute** should embrace the following 9-dimensional space of questions for consideration:

1. Where to transmit?
2. When to transmit?

3. At what wavelength?
4. What polarization to use?
5. What should be the energy of transmitting radio signal?
6. What modulation to apply?
7. What is the optimum structure for transmitted messages?
8. Why should we transmit interstellar radio messages?
9. What are the dangers of pursuing METI?

We shall try to give answers to all of the above questions. It is important to note that any such answers will be not final, but only preliminary in nature. As we have already emphasized, METI is a new, emerging human activity, and nothing that it implies is yet settled. Therefore, readers have a rare opportunity to join in discussions leading to a new scientific endeavor.

**1) Where to send interstellar radio messages?**

It has become much easier to answer this question since 1995, when an outstanding discovery was made. Swiss astronomer Michael Mayor and graduate student Didier Queloz announced in that year the detection of the first planet orbiting another Sun-like star, 51 Pegasus. Subsequent discoveries of well over 100 other exoplanets have made it clear that planets are ordinary celestial objects, as widespread as stars and galaxies. In our Galaxy alone, with on the order of 100 billion stars, 1% of them are stars of solar or nearly solar types. Here, among this remarkable billion, it is plausible to select stars to which our interstellar radio messages can be addressees. We do not propose restricting our targets to only these stars, but they should be our main goal, defined by our present understanding, recognizing that the question of other life sites is not yet settled, and that there remains an opportunity for further creativity and research. Our present list of requirements for candidate stars includes the following characteristics:

- Main sequence stars;
- Constant luminosity;
- Age in the range of 4 to 7 billion years;
- Single stars of spectral classes close to that of the Sun are preferable;
- Position in the sky close to "preferable directions" – near the ecliptic plane, in the direction of remarkable astronomical objects, toward the center or the anti-center of the Galaxy, etc.;
- It is desirable that we fall in the direction of remarkable astronomical objects as viewed from There, so that They might find us in the course of Their usual astronomical observations;
- In case of targets representing known planetary systems, it is desirable that orbits of these exoplanets have low eccentricity, as such planetary systems are more stable, and there is no significant temperature fluctuation interfering with the origin of life;
- It is desirable to choose stars inside the "Belt of a Life" – that "hothouse" area of our Galaxy, where because of coincidence of speeds of movement of stars and spiral sleeves, conditions for origin and long development of a life are believed optimum.

In due course, in the process of accumulating knowledge about the Cosmos, other criteria, and other locations than the stars addressed here, may emerge. For now, we propose concentrating on the above criteria.

**2) When to send IRMs to the selected star?**

Questions of time synchronization between our transmission and Their searches (or, in the case of SETI, between Their transmission and our searches) are very important. By Peter Makovetsky's estimation, as reported in his book *"Look in the Root"* ("Science" Publishing House, Moscow, 1979), competent synchronization allows us to increase the probability of establishing radio contact by a factor of tens. One possible method is to bind the moments of transmission ("Here") and searching ("There") to some well-known universal event which is observable everywhere in our Galaxy. For example, we could synchronize to the moment of maximum intensity of such explosions as Novae or Supernovae. Proceeding from simple geometrical parities, Makovetsky has calculated "schedules" for some neighboring stars in the case of we and They carrying out search coordinated to a Nova in the constellation, Cygnus which was observed on Earth on August, $29^{th}$, 1975. Using modern, large optical telescopes, it is now possible to register the moments of flashes of Supernovae in neighboring galaxies. These can also be used for time synchronization of messaging and searching in deep space.

3) **At what wavelength?**

The frequency band in which it is necessary to transmit IRMs coincides with that band which earlier has been proved most suitable for SETI – from 20 cm up to 1 cm, where the greatest range of radio communications is achieved. We define the energy potential of a space radio link as the product of power of the transmitter and the gain of the transmitting and receiving antennas, divided by the noise temperature of the receiving system. At current the state of development of our terrestrial technology, this relation is maximal in a centimetric band. We do not dismiss the possibility that, in due course, in the development of space communications, suitable energy potentials will be reached at infrared or optical wavelengths. Should that occur, our representations about optimum wavelength will of course change. Exact values of wavelength may even take on "magic" values. For example, 6.72 cm = 21 cm / Pi, would be known to all technological civilizations as the ratio of two universal constants, one physical (the radio emission line of interstellar neutral hydrogen) and the other mathematical.

4) **What polarization to use?**

The polarization integrity of a radiated signal is one possible indicator of artificial origin. In addition, by using polarization modulation the direction of rotation of circular polarization, or the orientation of the plane of linear polarization, can be varied discretely or continuously, as a means of encoding an intelligent message.

**5) What should be the energy of the transmitted radio signal?**

In the case of determining appropriate levels of power for dedicated transmitters specifically designed for continuous and systematic METI transmissions, estimates are readily computed. As for the somewhat different question of conducting METI now, using those instruments which currently exist, or will become available in the foreseeable future, a more important issue is the question not of transmitter power, but rather of realistic data rates to transmit meaningful information. The following summary shows computed data rates for METI experiments using the three most powerful transmitting radio systems now existing on Earth. The numbers in parentheses represent the diameter of the transmitting antenna, average power, and wavelength, respectively):

1. Radar Telescope in Arecibo, Puerto Rico (300 m; 1000 kW; 12.5 cm) – 1000 bits per second;
2. Solar System Planetary Radar in Goldstone, California (70 m; 480 kW; 3.5 cm) – 550 bits per second;
3. Planetary Radar near Evpatoria, Crimea (70 m; 150 kW; 6.0 cm) – 60 bits per second.

In these calculations, we assume the distance to which it is necessary to send our message is on the order of 70 light years, and we further assume that Their receiving system has the antenna with an effective aperture in 1 million square meters. A project to deploy just such a large radio astronomical antenna, the

Square Kilometer Array (SKA), is now under development on Earth, and could be constructed within the next decade.

## 6) What modulation to apply?

After more than 45 years of nearly continuous searches for intelligent signals from other civilizations, the overwhelming majority of studies employ surprisingly similar detection algorithms. It is accepted practice to apply digital spectral analysis, with the number of parallel analysis channels reaching hundreds of millions, and even several billions. For example, in its project "Phoenix" targeted search, the SETI Institute used a digital spectral analyzer of two million channels, with bin widths on the order of ~1 Hz. That allowed them to analyze, in real time, a bandwidth on the order of 2 MHz and on the order of 2 GHz in off-line mode!

Having assumed what exactly the optimum receiver should look for, not only in searching for radio signals from Other civilizations, but also in terms of such signals as we might transmit to ETIs, we come to the conclusion that such modulation should have a clear spectral signature, allowing decoding with minimal ambiguity, by means of the above-mentioned parallel spectral analyzers. One such modulation format, well known and widely used on Earth, is frequency modulation (FM).

## 7) What are the optimum structure and method of encoding a transmitted message?

Having suggested that a radio message should be synthesized on the basis of self-evident and physically proven spectral constituents, we now propose the following structure (Table 1). We identify three types of single-valued frequency function: "Constant", "Continuous", and "Discrete." A radio message to ETI message could employ a three-section structure and incorporates three specific languages, which we can call "the language of nature", "the language of emotions", and "the language of logic". In Table 1 the term "Sonogram" designates a visualization of the spectral structure of a signal in Cartesian coordinates [axis $X$ = frequency, axis $Y$ = time].

### Table1. Spectral Language for Messaging to ETI

| Parameter | Three types of single-valued frequency function | | |
|---|---|---|---|
| Type | 1. Constant | 2. Continuous | 3. Discrete |
| Author (*"Here"*) | Radio Engineer | Composer, Painter, Architect | Scientist |
| Language | "of Nature" | "of Emotion" | "of Logic" |
| Information | Is absent | Analog | Digital |
| Sonogram of transmitting signals | 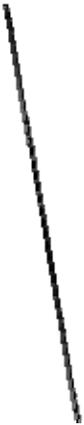 | 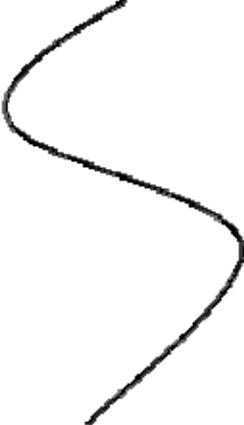 | 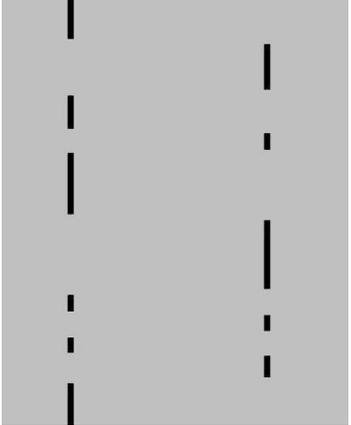 |
| Analysis (*"There"*) | Astrophysical | Art criticism | Linguistic |

Here, we draw a pertinent analogy with a suggested triune structure of human thinking, wherein we distinguish three components – *intuitive, emotional, and logical.* The first part of such a three-part radio message is designed by radio engineers, and represents a coherent signal, for example, elementary monochromatic CW or periodic LFM (linear frequency modulation). It is possible to adjust its frequency for variable Doppler correction, such that our modulated signal will be observed by aliens as a constant frequency. We suggest that ETI will intuitively understand the significance of a sounding signal thus sent. The second part of the message is created by people versed in the arts – composers, artists, architects – and consists of analog variations of frequency, representing our emotional world and our artistic conceptions. An elementary example is classical musical melodies. The third part of our message consists of discrete frequency shift keying, a digital dataflow, representing our logic constructions – algorithms, theories, cumulative knowledge about us and about the world around us. In the line "Analysis" our representations are displayed in terms of how such signals will be investigated "There", on the reception side (or "Here", in case of success of terrestrial SETI). The first message part is optimized for astrophysical analysis, with the purpose of revealing the effects of the interstellar environment, and supporting diagnostics of a propagation channel. The second part is analyzed by art critics; the third, by linguists, logicians, and other scientists.

**8) Why should we transmit interstellar radio messages?**

Here, we step onto the shaky ground of "fuzzy and imprecise" reasoning and assumptions. A strict proof of the necessity and practicality of METI is of course impossible. Emotional and ethical reasons of a messianic and altruistic nature, such as "to bring to Aliens a long-awaited message that they are not alone in the Universe", are convincing and inspirational to only the few. Nevertheless, we should understand a simple thing – if all civilizations in the Universe are only recipients, and not message-sending civilizations, than no SETI searches make any sense…

**9) Is it dangerous to engage in METI?**

We can refer to a fear of transmitting from Earth as ***METI-phobia***. It has its roots in fears expressed right after transmission of the first interstellar radio message, from Arecibo in 1974. The Nobel laureate Martin Ryle, a prominent radio astronomer, publicly proposed imposing an interdiction of any attempts at messaging from Earth to prospective extraterrestrial civilizations.

Our understanding of the given problem starts with a certain "double standard": many vocal and impressionable people are afraid of a super-powerful and super-aggressive "Something" from which there is no salvation, and which has either already long ago found us, or which will by all means soon find us, from the radio emission of tens of powerful military radars in the USA and Russia, which formed the basis of their national ballistic missile warning systems, working continuously, day and night, since the sixties of the last century. Thus, even in the case of civilizations as primitive and power limited as we, detection over prodigious distances can already be assumed.

**Realized METI projects**

Throughout the entire history of our civilization, only four projects involving transmitting of interstellar radio messages (IRMs) have yet been fully developed and realized. In Table 2, these four projects are ordered by the dates of the first transmitting sessions (in total, as it can be concluded from the table, only 16 such sessions have ever taken place). The symbols T and E here represent the total transmit duration in minutes, and radiated energy in Mega Joules, of each of the four METI projects conducted to date.

Table 2. Realized METI projects

| Name | Arecibo Message | Cosmic Call 1 | Teen Age Message | Cosmic Call 2 |
|---|---|---|---|---|
| Date | 16.11.1974 | 24.05, 30.06, 01.07.1999 | 29.08, 03.09, 04.09.2001 | 06.07.2003 |
| Type | First IRM (digital) | First multi page IRM | First analog and digital IRM | First international IRM |

| Authors | Drake, Sagan, Issacman, et al | Chafer, Dutil, Dumas, Braastad, Zaitsev, et al | Pshenichner, Gindilis, Zaitsev, et al. | Chafer, Dutil, Dumas, Braastad, Zaitsev, et al |
|---|---|---|---|---|
| Radar | Arecibo | Evpatoria | Evpatoria | Evpatoria |
| Sessions | 1 | 4 | 6 | 5 |
| T, min | 3 | 960 | 366 | 900 |
| E, MJ | 83 | 8640 | 2200 | 8100 |
| Ref. | [1] | [2] | [3] | [4] |

The noteworthy Arecibo Message of 1974 [1] had the size of 1679 bits, and was sent to globular cluster M13. It is extensively described in the literature and on the Internet; therefore we will not further elaborate on it.

25 years later, transmitting of interstellar radio messages was renewed, using the Evpatoria planetary radar. In 1999, the "Cosmic Call 1" IRM [2] was transmitted to 4 Sun-like stars. It represented a peculiar encyclopedia, including a terrestrial overview about us and the world around us, written in a special language called Lexicon, as well as data about the "Cosmic Call" project and its participants. In structure, Cosmic Call 1 closely paralleled the Arecibo Message. The size of this "Encyclopedia" was 370967 bits.

In 2001 the "Teen Age Message" [3] was sent to 6 Sun-like stars. Here is the first and, unfortunately, so far the only time the three-section structure described above has been applied – a monochromatic sounding signal was first radiated, then the analog information (music), and finally, a digital message was transmitted. As a source for the analog portion of the transmission, quasi-monochromatic signals with a low level of overtones from the "Theremin" electric musical instrument were included. Such a signal greatly facilitates detection and perception over interstellar distances. The digital part of the message consisted of 28 binary, Arecibo-like, images with a total size of 648220 bits.

In 2003 IRM "Cosmic Call 2" [4] was sent to 5 Sun-like stars. This was the first international IRM, and fragments of all three previous radio messages were included in it. We consider that all future messages from the Earth should have precisely such international content.

Table 3 summarizes the expected times at which the Arecibo and the three Evpatoria messages sent to date will arrive at their corresponding target stars. The second column of Table 3 predicts the time when the era of the "Great Silence of the Universe" can potentially end for those at the receiving side of the communications link, in the optimistic event that They are there, and given the "happy case" that they should happen to find these particular intelligent signals from our terrestrial Civilization.

Table 3. Arrival Time of Terrestrial Radio Messages

| # | Arrival | Star | Constellation | Message | Sent, d m y | Distance, l.y. |
|---|---|---|---|---|---|---|
| 1 | Apr 2036 | Hip 4872 | Cassiopeia | Cosmic Call 2 | 06.07.2003 | 32.8 |
| 2 | Aug 2040 | HD 245409 | Orion | Cosmic Call 2 | 06.07.2003 | 37.1 |
| 3 | May 2044 | HD 75732 | Cancer | Cosmic Call 2 | 06.07.2003 | 40.9 |
| 4 | Sep 2044 | HD 10307 | Andromeda | Cosmic Call 2 | 06.07.2003 | 41.2 |
| 5 | Jul 2047 | HD 95128 | Ursa Major | Teen Age Message | 03.09.2001 | 45.9 |
| 6 | May 2049 | HD 95128 | Ursa Major | Cosmic Call 2 | 06.07.2003 | 45.9 |
| 7 | Apr 2051 | HD 190360 | Cygnus | Cosmic Call 1 | 01.07.1999 | 51.8 |
| 8 | Feb 2057 | HD 190406 | Sagitta | Cosmic Call 1 | 30.06.1999 | 57.6 |

| 9 | May 2057 | HD 76151 | Hydra | Teen Age Message | 04.09.2001 | 55.7 |
| 10 | Dec 2057 | HD 50692 | Gemini | Teen Age Message | 03.09.2001 | 56.3 |
| 11 | Jan 2059 | HD 126053 | Virgo | Teen Age Message | 03.09.2001 | 57.4 |
| 12 | Jan 2059 | HD 193664 | Draco | Teen Age Message | 04.09.2001 | 57.4 |
| 13 | Oct 2067 | HD 178428 | Sagitta | Cosmic Call 1 | 30.06.1999 | 68.3 |
| 14 | Nov 2069 | HD 186408 | Cygnus | Cosmic Call 1 | 24.05.1999 | 70.5 |
| 15 | Feb 2070 | HD 197076 | Delphinus | Teen Age Message | 29.08.2001 | 68.5 |
| 16 | ~25974 | Glob cluster M13 | Hercules | Arecibo | 16.11.1974 | ~24000 |

As a matter of fact, at moment of such detection, They can be said to begin living in an inhabited Universe. This revolution, not only in Their consciousness, but also in the Universe as a whole, can potentially be made by *us* – by our intellect and our good will. In truth, this is the most worthy application of terrestrial Reason!

***Acknowledgments.*** I thank Paul Shuch for his hard work on correction of my imperfect English and for his useful remarks. I am grateful to Jon Lomberg, Donald Goldsmith, Seth Shostak, Michael Matessa, and Richard Braastad for their useful comments on the manuscript.

See also web pages "Interstellar Radio Messages"

<http://www.cplire.ru/html/ra&sr/irm/index.html> (in English)

<http://www.cplire.ru/rus/ra&sr/index.html> (in Russian).